\documentclass[aps,prl,reprint,superscriptaddress]{revtex4-2}
\usepackage{graphicx}
\usepackage{braket}
\usepackage{amsmath}
\usepackage{amsthm}
\usepackage{amssymb}
\usepackage[colorlinks,
            linkcolor=blue,
            anchorcolor=black,
            citecolor=blue,
			urlcolor=blue]{hyperref}

\usepackage{algorithm}  
\usepackage{algorithmic}  
\usepackage{makecell}
\usepackage{array}
\usepackage{color}

\newtheorem{proposition}{Proposition}



\begin{document}


\title{Recursive expansion of Tanner graph: a method to construct stabilizer codes with high coding rate}


\author{Zhengzhong Yi}
\author{Zhipeng Liang} 
\author{Zicheng Wang}
\author{Jiahan Chen}
\author{Chen Qiu}
\author{Yulin Wu}
\author{Xuan Wang}
\email[]{wangxuan@cs.hitsz.edu.cn}
\affiliation{Harbin Institute of Technology, Shenzhen. Shenzhen, 518055, China}


\date{\today}

\begin{abstract}
	Quantum stabilizer codes face the problem of low coding rate. In this article, following the idea of recursively expanding Tanner graph proposed in our previous work, we try to construct new stabilizer codes with high coding rate, and propose XZ-type Tanner-graph-recursive-expansion (XZ-TGRE) code and Tanner-graph-recursive-expansion hypergraph product (TGRE-HP) code. XZ-TGRE code have zero asymptotic coding rate, but its coding rate tends to zero extremely slowly with the growth of code length. Under the same code length, its coding rate is much higher than that of surface code. The coding rate of TGRE-HP is the constant 0.2, which is the highest constant coding rate of stabilizer codes to our best knowledge. We prove that the code distance of XZ-TGRE code scales as $O(log(N))$, and that of TGRE-HP code scales as $O(\log \sqrt{N})$, where $N$ is the code length. Moreover, the code capacity noise threshold of XZ-TGRE code is around 0.078, and that of TGRE-HP code is around 0.096. This articles shows that the idea of recursively expanding Tanner graph might have potential to construct quantum codes with good performance. 

\end{abstract}


\maketitle

\section{Introduction}
\label{introduction}
Quantum error correcting codes provide us with a promising approach to achieving large-scale universal fault-tolerant quantum computing. However, most quantum error correcting codes face the problem of low coding rate, such as surface code\cite{bravyi1998quantum,fowler2012surface}, XZZX surface code\cite{bonilla2021xzzx}, concatenated code\cite{knill2005quantum}, 3D toric code\cite{kubica2019cellular}, and 4D toric code\cite{breuckmann2016local}, whose asymptotic coding rate are all zero. (3,4) hypergraph product code\cite{grospellier2021combining}, (5,6) hypergraph product code\cite{grospellier2018numerical} and  (4,5)-hyperbolic
surface code\cite{breuckmann2016constructions} are reported to have constant coding rate. However, their coding rates are 0.04, 0.016 and 0.1 respectively, which are still low compared to classical error correcting codes, such as polar code\cite{arikan2009channel} and LDPC code\cite{ryan2004introduction}. In quantum error correcting, to obtain lower logical error rate after error correcting, one often need to use longer code. In this situation, low coding rate will cause substantial physical qubit overhead. 

In our previous work \cite{yi2022quantum}, we propose a new idea to design stabilizer codes by recursively expanding Tanner graph, and a new stabilizer code named Z-type Tanner-graph-recursive-expansion (Z-TGRE) code (we haven't name them in \cite{yi2022quantum}, here we formally name them). Z-TGRE code have constant coding rate of 0.5, but can only correct Pauli $X$ and $Y$ errors. By simply replacing the Pauli $Z$ of Z-TGRE code by Pauli $X$ or $Y$, one can obtain X-TGRE code or Y-TGRE code which can correct Pauli $Y$ and $Z$ errors, or Pauli $X$ and $Z$ errors. 

In this article, following the idea of recursively expanding Tanner graph, we propose a new stabilizer code called XZ-type Tanner-graph-recursive-expansion (XZ-TGRE) code, which can correct all three Pauli errors. Though XZ-TGRE code still have zero asymptotic coding rate, its coding rate tends to zero extremely slowly. In a fairly long code length range, its coding rate is higher than some stabilizer codes with constant coding rate. We prove that the code distance of XZ-type Tanner-graph-recursive-expansion code scales as $O(\log (N))$, where $N$ is the code length. 

Moreover, using the idea proposed by Tillich and Zémor\cite{tillich2013quantum}, through performing Cartesian product of the Tanner graphs of X-TGRE code and Z-TGRE code, we propose a new class of hypergraph product codes named Tanner-graph-recursive-expansion hypergraph product (TGRE-HP) code, whose code distance scales as $O(\log \sqrt{N})$ and coding rate is the constant 0.2. To our best knowledge, this coding rate is the highest among the existing quantum stabilizer codes.

Through simulation with fully-decoupled belief propagation (FDBP) decoding algorithm\cite{yi2023improved}, we find the code capacity noise threshold\cite{landahl2011fault} is around 0.078, and that of TGRE-HP code is around 0.096. The code capacity threshold of these two codes is much higher than that of 4D-hyperbolic code\cite{breuckmann2021single} (around 0.050), whose coding rate is the constant 0.18, and that of (4,5)-hyperbolic surface code\cite{breuckmann2016constructions} (around 0.025), whose coding rate is the constant 0.1. It should be noticed that the weight of stabilizer of both the XZ-TGRE code and TGRE-HP code will grow with the increase of code length. Since the decoding accuracy of FDBP will be reduced by the increase of the weight of stabilizers, the actual code capacity of these two codes might even be higher than our simulation results. 


\section {Z-TGRE, X-TGRE and Y-TGRE code}
\label{Z-TGRE code}

\begin{figure}[htbp]
	\centering
	\includegraphics[width=0.48\textwidth]{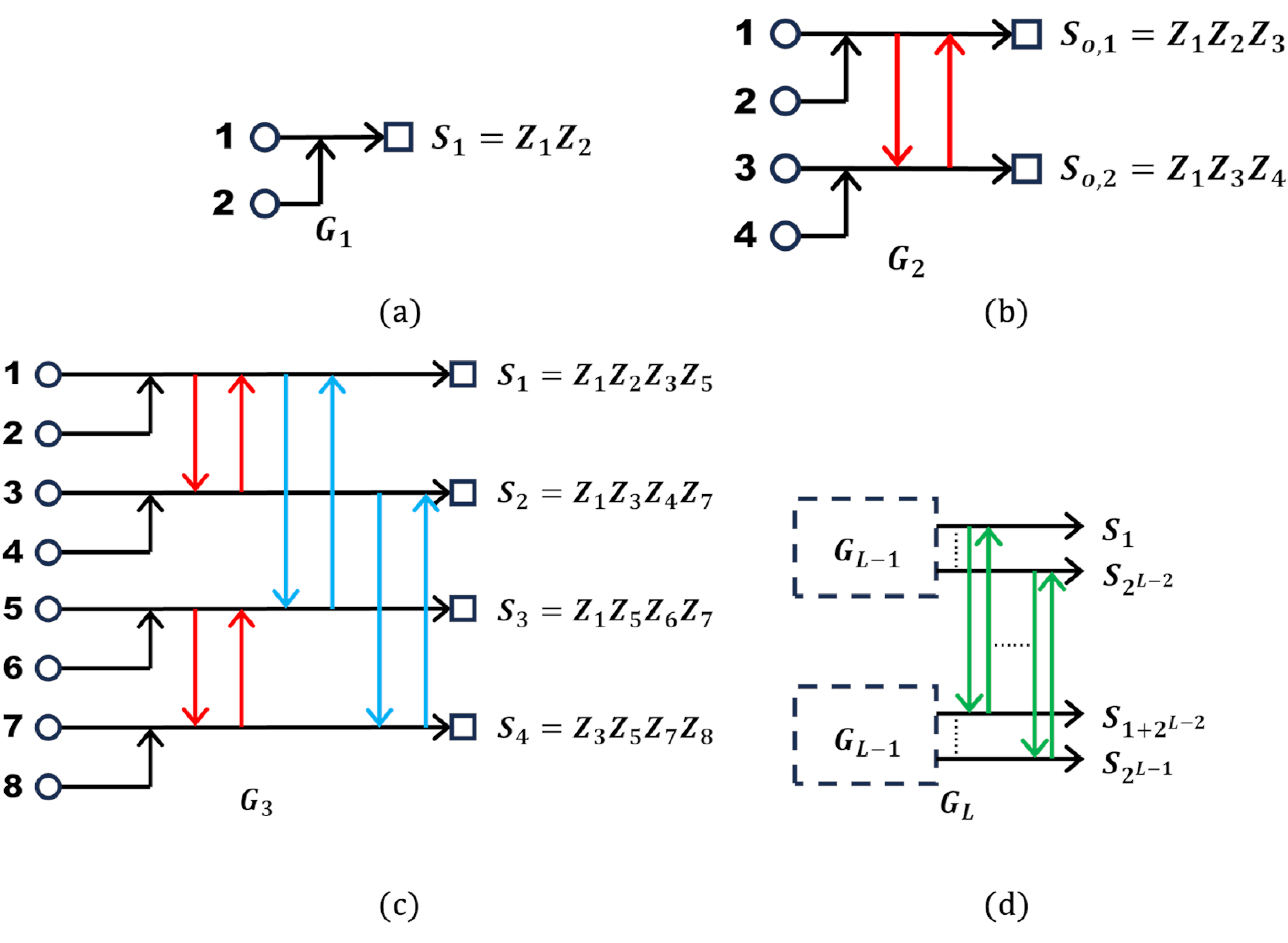}
	\caption{The recursive expansion of Tanner graph of Z-TGRE code. The arrows mean the corresponding variable nodes they starts from will join in the corresponding check nodes they ends with. The variable nodes are numbered from 1 to $N=2^L$. $S_i$ denotes the $i$th stabilizer generated by the corresponding check nodes in the graphs. (a) The primal Tanner graph $G_1$. (b) Tanner graph $G_2$ by the expansion of two primal Tanner graphs $G_1$. (c) Tanner graph $G_3$ by the expansion of two $G_2$. (d) Tanner graph $G_L$ by the expansion of two $G_{L-1}$.}
	\label{Z-TGRE Tanner graph}
\end{figure}

Z-TGRE code, which can correct Pauli $X$ and $Y$ errors, is constructed through recursively expanding Tanner graph. The way to expand Tanner graph is shown in Fig. \ref{Z-TGRE Tanner graph}. By simply replacing the Pauli $Z$ of Z-TGRE code by Pauli $X$ or $Y$, one can obtain X-TGRE code or Y-TGRE code which can correct Pauli $Y$ and $Z$ errors, or Pauli $X$ and $Z$ errors. The independence and commutativity of the stabilizers obtained from Fig. \ref{Z-TGRE Tanner graph} have been explained in \cite{yi2022quantum}. It's obvious when the code length is $N$, the number of the stabilizers and the number of logical qubits are both $N/2$. Hence, the coding rate is the constant 0.5. 

As shown in Table \ref{table_Z-TGRE}, logical $X$ operator $\bar{X}_i$ consists of several Pauli $X$s, and logical $Z$ operators $\bar{Z}_i$ consists of single Pauli $Z$. If $L$ is odd, the subscripts of the expression $\bar{X}_i$ are the same as those of stabilizer $S_i$'s. If $L$ is even, the subscripts of the expression of $\bar{X}_i$ can be obtained by subtracting 1 from the even subscript and adding 1 to all odd subscripts of the expression of $\bar{S}_i$. As for logical $Z$ operators, the subscript of the expression of $\bar{Z}_i$ is the unique number in those of $\bar{X}_i$. 

\begin{table*}[htbp]
	\begin{center}
		\caption{The stabilizer generators and corresponding logical operators of Z-TGRE code with code length $N=4, 8, 16, 32$.}
		\label{table_Z-TGRE}
		\begin{tabular}{c|c|c|c}
			\textbf{Code length} & \ \textbf{Stabilizers} & \ \textbf{Logical X operators} & \ \textbf{Logical Z operators}\\
			\hline
			\rule{0pt}{10pt}{4} & $S_1=Z_{1}Z_{2}Z_{3}$ & $\bar{X}_1=X_{1}X_{2}X_{4}$ & $\bar{Z}_1=Z_{1}$ \\
			& $S_2=Z_{1}Z_{3}Z_{4}$ & $\bar{X}_2=X_{2}X_{3}X_{4}$ & $\bar{Z}_2=Z_{3}$\\
			\hline
			\rule{0pt}{10pt}{8} & $S_1=Z_{1}Z_{2}Z_{3}Z_{5}$ & $\bar{X}_1=X_{1}X_{2}X_{3}X_{5}$ & $\bar{Z}_1=Z_{2}$ \\
			& $S_2=Z_{1}Z_{3}Z_{4}Z_{7}$ & $\bar{X}_2=X_{1}X_{3}X_{4}X_{7}$ & $\bar{Z}_2=Z_{4}$ \\
			& $S_3=Z_{1}Z_{5}Z_{6}Z_{7}$ & $\bar{X}_3=X_{1}X_{5}X_{6}X_{7}$ & $\bar{Z}_3=Z_{6}$ \\
			& $S_4=Z_{3}Z_{5}Z_{7}Z_{8}$ & $\bar{X}_4=X_{3}X_{5}X_{7}X_{8}$ & $\bar{Z}_4=Z_{8}$ \\
			\hline
			\rule{0pt}{10pt}{16} & $S_1=Z_{1}Z_{2}Z_{3}Z_{5}Z_{9}$ & $\bar{X}_1=X_{1}X_{2}X_{4}X_{6}X_{10}$ & $\bar{Z}_1=Z_{1}$ \\
			& $S_2=Z_{1}Z_{3}Z_{4}Z_{7}Z_{11}$ & $\bar{X}_2=X_{2}X_{3}X_{4}X_{8}X_{12}$ & $\bar{Z}_2=Z_{3}$ \\
			& $S_3=Z_{1}Z_{5}Z_{6}Z_{7}Z_{13}$ & $\bar{X}_3=X_{2}X_{5}X_{6}X_{8}X_{14}$ & $\bar{Z}_3=Z_{5}$ \\
			& $S_4=Z_{3}Z_{5}Z_{7}Z_{8}Z_{15}$ & $\bar{X}_4=X_{4}X_{6}X_{7}X_{8}X_{16}$ & $\bar{Z}_4=Z_{7}$ \\
			& $S_5=Z_{1}Z_{9}Z_{10}Z_{11}Z_{13}$ & $\bar{X}_5=X_{2}X_{9}X_{10}X_{12}X_{14}$ & $\bar{Z}_5=Z_{9}$ \\
			& $S_6=Z_{3}Z_{9}Z_{11}Z_{12}Z_{15}$ & $\bar{X}_6=X_{4}X_{10}X_{11}X_{12}X_{16}$ & $\bar{Z}_6=Z_{11}$ \\
			& $S_7=Z_{5}Z_{9}Z_{13}Z_{14}Z_{15}$ & $\bar{X}_7=X_{6}X_{10}X_{13}X_{14}X_{16}$ & $\bar{Z}_7=Z_{13}$ \\
			& $S_8=Z_{7}Z_{11}Z_{13}Z_{15}Z_{16}$ & $\bar{X}_8=X_{8}X_{12}X_{14}X_{15}X_{16}$ & $\bar{Z}_8=Z_{15}$ \\
			\hline
			\rule{0pt}{10pt}{32} & $S_1=Z_{1}Z_{2}Z_{3}Z_{5}Z_{9}Z_{17}$ & $\bar{X}_1=X_{1}X_{2}X_{3}X_{5}X_{9}X_{17}$ & $\bar{Z}_{1}=Z_{2}$ \\
			& $S_2=Z_{1}Z_{3}Z_{4}Z_{7}Z_{11}Z_{19}$ & $\bar{X}_2=X_{1}X_{3}X_{4}X_{7}X_{11}X_{19}$& $\bar{Z}_{2}=Z_{4}$\\
			& $S_3=Z_{1}Z_{5}Z_{6}Z_{7}Z_{13}Z_{21}$ & $\bar{X}_3=X_{1}X_{5}X_{6}X_{7}X_{13}X_{21}$& $\bar{Z}_{3}=Z_{6}$\\
			& $S_4=Z_{3}Z_{5}Z_{7}Z_{8}Z_{15}Z_{23}$ & $\bar{X}_4=X_{3}X_{5}X_{7}X_{8}X_{15}X_{23}$& $\bar{Z}_{4}=Z_{8}$\\
			& $S_5=Z_{1}Z_{9}Z_{10}Z_{11}Z_{13}Z_{25}$ & $\bar{X}_5=X_{1}X_{9}X_{10}X_{11}X_{13}X_{25}$& $\bar{Z}_{5}=Z_{10}$\\
			& $S_6=Z_{3}Z_{9}Z_{11}Z_{12}Z_{15}Z_{27}$ & $\bar{X}_6=X_{3}X_{9}X_{11}X_{12}X_{15}X_{27}$& $\bar{Z}_{6}=Z_{12}$\\
			& $S_7=Z_{5}Z_{9}Z_{13}Z_{14}Z_{15}Z_{29}$ & $\bar{X}_7=X_{5}X_{9}X_{13}X_{14}X_{15}X_{29}$
			& $\bar{Z}_{7}=Z_{14}$\\
			& $S_8=Z_{7}Z_{11}Z_{13}Z_{15}Z_{16}Z_{31}$ & $\bar{X}_8=X_{7}X_{11}X_{13}X_{15}X_{16}X_{31}$& $\bar{Z}_{8}=Z_{16}$\\
			& $S_9=Z_{1}Z_{17}Z_{18}Z_{19}Z_{21}Z_{25}$ & $\bar{X}_9=X_{1}X_{17}X_{18}X_{19}X_{21}X_{25}$& $\bar{Z}_{9}=Z_{18}$\\
			& $S_{10}=Z_{3}Z_{17}Z_{19}Z_{20}Z_{23}Z_{27}$ & $\bar{X}_{10}=X_{3}X_{17}X_{19}X_{20}X_{23}X_{27}$& $\bar{Z}_{10}=Z_{20}$\\
			& $S_{11}=Z_{5}Z_{17}Z_{21}Z_{22}Z_{23}Z_{29}$ & $\bar{X}_{11}=X_{5}X_{17}X_{21}X_{22}X_{23}X_{29}$& $\bar{Z}_{11}=Z_{22}$\\
			& $S_{12}=Z_{7}Z_{19}Z_{21}Z_{23}Z_{24}Z_{31}$ & $\bar{X}_{12}=X_{7}X_{19}X_{21}X_{23}X_{24}X_{31}$& $\bar{Z}_{12}=Z_{24}$\\
			& $S_{13}=Z_{9}Z_{17}Z_{25}Z_{26}Z_{27}Z_{29}$ & $\bar{X}_{13}=X_{9}X_{17}X_{25}X_{26}X_{27}X_{29}$& $\bar{Z}_{13}=Z_{26}$\\
			& $S_{14}=Z_{11}Z_{19}Z_{25}Z_{27}Z_{28}Z_{31}$ & $\bar{X}_{14}=X_{11}X_{19}X_{25}X_{27}X_{28}X_{31}$& $\bar{Z}_{14}=Z_{28}$\\
			& $S_{15}=Z_{13}Z_{21}Z_{25}Z_{29}Z_{30}Z_{31}$ & $\bar{X}_{15}=X_{13}X_{21}X_{25}X_{29}X_{30}X_{31}$& $\bar{Z}_{15}=Z_{30}$\\
			& $S_{16}=Z_{15}Z_{23}Z_{27}Z_{29}Z_{31}Z_{32}$ & $\bar{X}_{16}=X_{15}X_{23}X_{27}X_{29}X_{31}X_{32}$& $\bar{Z}_{16}=Z_{32}$\\
			\hline
		\end{tabular}
		
	\end{center}
	
\end{table*}

The capability to correct Pauli $X$ and $Y$ errors depends on the minimum weight of logical $Z$ operators, which is given in Proposition \ref{the minimum weight of logical X of Z-TGRE}.

\begin{proposition}
	\label{the minimum weight of logical X of Z-TGRE}
	For Z-TGRE code with code length $N=2^L$, if $L$ is even, the minimum weight of logical operators $\bar{X}$ is $\log N$. If $L$ is odd, the minimum weight of logical operators $\bar{X}$ is $\log N+1$.
\end{proposition} 

\begin{figure}
	\centering
	\includegraphics[width=0.5\textwidth]{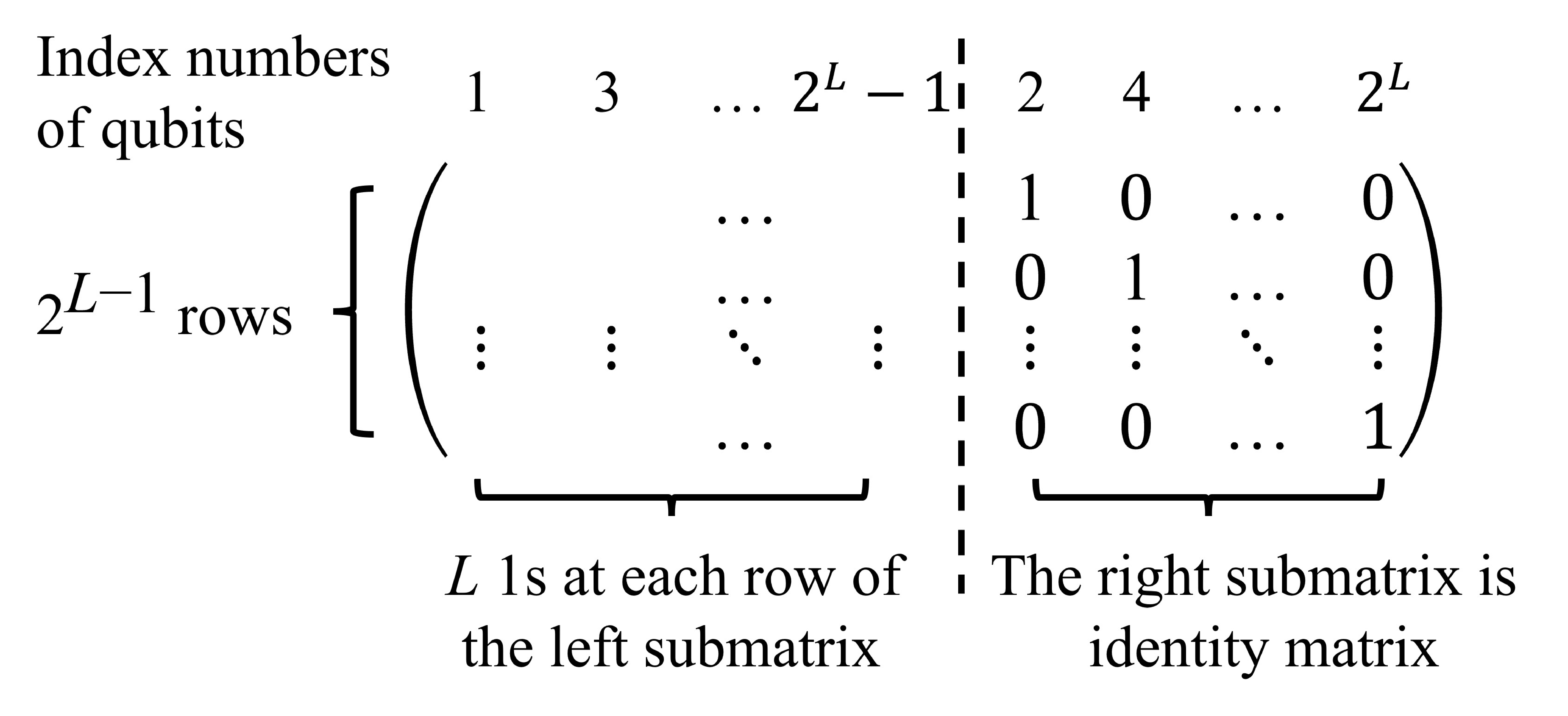}
	\caption{The structure of $H_L$.}
	\label{The structure of Hl}
\end{figure}

\begin{figure*}
	\centering
	\includegraphics[width=1\textwidth]{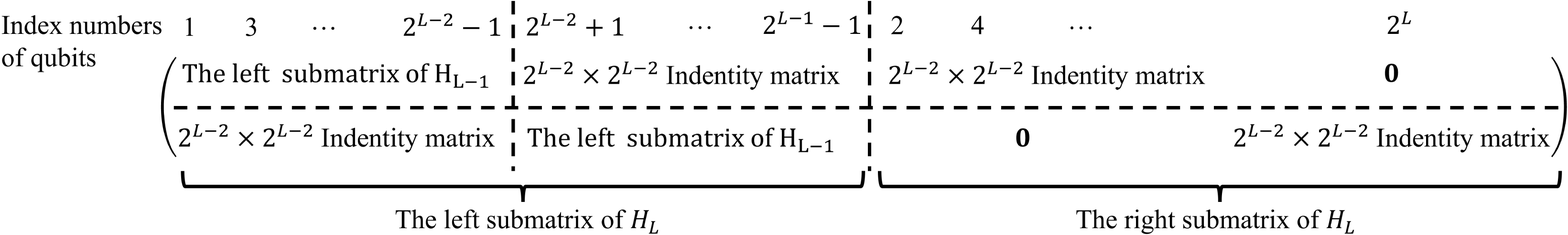}
	\caption{The recursive structure of $H_L$.}
	\label{The recursive structure of Hl}
\end{figure*}

\begin{proof}
	The whole proof is divided into 4 steps:
	
	(1) Transforming the problem of computing the minimum weight of logical $X$ operators to the problem of computing the minimum weight of stabilizers.
	
	(2) Transforming the problem of computing the minimum weight of stabilizers to the problem of computing the minimum weight of a certain (the first) stabilizer.
	
	(3) Giving the expansion method of parity check matrix corresponding to the recursive expansion of the Tanner graph as shows in Fig. 1 in the main text.
	
	(4)	For a given code length, employing mathematical induction to prove the minimum weight of the stabilizer by the expansion method of parity check matrix.
	
	The detailed process for each step of the proof is as follows:
	
	(1) For a Z-TGRE code with code length $N=2^L$, there is one-to-one relationship between its logical $X$ operators and stabilizer generators. Specifically, when $L$ is odd, logical $X$ operators can be obtained by replacing the Pauli $Z$ operators of stabilizer generators by Pauli $X$ operators. When $L$ is even, logical $X$ operators can be obtained by replacing the Pauli $Z$ operators of stabilizer generators by Pauli $X$ operators and subtracting all even subscripts by 1 (i.e. replacing even number $i$ by odd number $i-1$ as shown in Fig. 1 in the main text), while adding 1 to all odd subscripts (i.e. replacing the odd number $j$ by even number $j+1$ as shown in Fig. 1 in the main text). Thus, for a given code length, it is only need to prove the minimum weight of all stabilizers to prove the minimum weight of their corresponding logical $X$ operator.
	
	(2) Observing the recursive expansion as shown in Fig. 1 in the main text, it can be seen that exchanging the expression of stabilizers $S_i$ and $S_j$ can be realized by the permutation of the index numbers of qubits. Thus, we only need to prove the minimum weight of a stabilizer $S_i$ to prove the minimum weight of all stabilizers. In this proof, we prove the minimum weight of $S_1$.
	
	(3) Noticed that, according to the recursive expansion shown in Fig. 1 in the main text, when code length $N=2^L$, the parity-check matrix $H_L$ can be divided into two submatrices with the same number of columns on the left and right, where the left submatrix corresponds to qubits with odd index numbers and the right submatrix corresponds to qubits with even index numbers as shown in Fig. \ref{The structure of Hl}.

	The expansion method of parity check matrix corresponding to the recursive expansion of Tanner graph as shown in Fig. 1 in the main text is as follows: the parity-check matrix $H_L$ corresponding to code length $N=2^L$ is composed of $H_{L-1}$ corresponding to code length $N=2^{L-1}$ in the manner as shown in Fig. \ref{The recursive structure of Hl}.

	(4) Next, using mathematical induction to prove the minimum weight of stabilizer $S_1$. $S_1$ corresponds to the first row of parity check matrix. When $L=b=2$ and $L=b=3$, it is easy to prove that Proposition \ref{the minimum weight of logical X of Z-TGRE} holds. When $L=b=2$, the weight of the first row of the parity check matrix can be reduced to 2 by some two rows. When $L=b=3$, the weight of the first row of the parity check matrix is 4. When combining it with any other two rows (i.e. multiplying 3 rows in total), its weight is still 4.
	
	First, proving that if $L=b$ is even, Proposition \ref{the minimum weight of logical X of Z-TGRE} holds, and hence Proposition \ref{the minimum weight of logical X of Z-TGRE} also holds when $L=b+1$.
	
	Assuming that Proposition \ref{the minimum weight of logical X of Z-TGRE} holds when $L=b\geq2$, and multiplying the first row with some $r_b=b-1$ rows can reduce the weight to $b$.
	
	When $L=b+1$, we should preferentially select the first $2^{L-2}$ rows to reduce the weight of the first row of the parity-check matrix $H_L$. The reason is that the first row of $H_L$ contains the first row of the left submatrix of $H_{L-1}$ which contains $b$ 1s. If we use the latter $2^{L-2}$ rows of $H_L$ to reduce the weight of the first row, then at most one 1 in the first row belongs to the left submatrix of $H_{L-1}$ will be eliminated each time, and at the same time, one 1 will be introduced to the part of the first row belongs to the right submatrix of $H_L$. However, if we use the first $2^{L-2}$ rows to reduce the weight of the first row, two 1s in the first row of the left submatrix of $H_{L-1}$ can be eliminated at most each time, and one $1$ will be introduced to the part of the first row belongs to the right submatrix of $H_L$. Thus, we should preferentially select the first $2^{L-2}$ rows to reduce the weight of the first row of the parity-check matrix $H_L$. According to the case where $L=b$, by selecting $r_b=b-1$ rows from the first $2^{L-2}$ rows and multiplying them with the first row, the minimum weight of the first row can be reduced to $b+r_b+1$ (the weight of the part belongs to the left submatrix is $b$, and the weight of the part belongs to the right submatrix is $r_b+1$). At this point, at most one row can be selected from the latter $2^{L-2}$ rows to multiply with the first row to reduce its weight. Otherwise, the number of rows selected to reduce the weight of the first row is at least $b+1=L$, which will inevitably result that the part of the first row belongs to the right submatrix contains at least $L+1$ $1$s. If we select one row from the latter $2^{L-2}$ rows to reduce the weight, since each row of the left submatrix of $H_L$ contains $b+1$ (odd) $1$s, the part belongs to the left submatrix of the first row will be eliminated by up to one (even) $1$s (since its minimum weight is $b$), and one $1$ is introduced. At the same time, this operation will introduce one 1 to the part of the first row belongs to the right submatrix. The weight of the first row right now is
	\begin{equation}
		\label{wts11}
		wt\left(S_1\right)\geq b+r_b+1-b+2=L+1
	\end{equation}
	Observing the recursive expansion in Fig. 1 in the main text, it is easy to verify equality holds in Eq. (\ref{wts11}) when multiply the first row of $H_L$ with the $j$th $\left(j=1+2^0,1+2^1,\cdots,1+2^{L-2}\right)$ row.
	
	Second, proving that if $L=b$ is odd, Proposition \ref{the minimum weight of logical X of Z-TGRE} holds, and hence Proposition \ref{the minimum weight of logical X of Z-TGRE} also holds when $L=b+1$.
	
	Assuming that Proposition \ref{the minimum weight of logical X of Z-TGRE} holds when $L=b$ is odd, and multiplying the first row with some $r_b=b-1$ rows can reduce the weight to $b+1$.
	
	When $L=b+1$, as described above, we should still preferentially select the first $2^{L-2}$ rows to reduce the weight of the first row of the parity-check matrix $H_L$, based on which at most one row from the latter $2^{L-2}$ rows can be choose to multiply with the first row to reduce the weight. Since each row of the left submatrix of $H_L$ contains $b+1$ (even) $1$s, and the weight of the part of the first row belongs to the left submatrix is $b+1$, which might be eliminated by up to $b+1$ (even) $1$s and add one $1$ to the part of the first row belonging to the right submatrix. The weight of the first row right now is
	\begin{equation}
		\label{wts12}
		wt\left(S_1\right)\geq b+1+r_b+1-(b+1)+1=L
	\end{equation}
	
	Observing the recursive method in Fig. 1 in the main text, it is easy to verify that the weight of the first row is $L$ and equality holds in Eq. (\ref{wts12}). When multiplying the first row of $H_L$ with the jth $\left(j=1+2^0,1+2^1,\cdots,1+2^{L-2}\right)$ row, the equality also holds in Eq. (\ref{wts12}).
	
	The proof is completed.
\end{proof}

The capability to correct Pauli $X$ and $Y$ errors has been verified in \cite{yi2022quantum}.

\section {XZ-TGRE code}
\label{XZ-TGRE code}

\begin{figure}[htbp]
	\centering
	\includegraphics[width=0.3\textwidth]{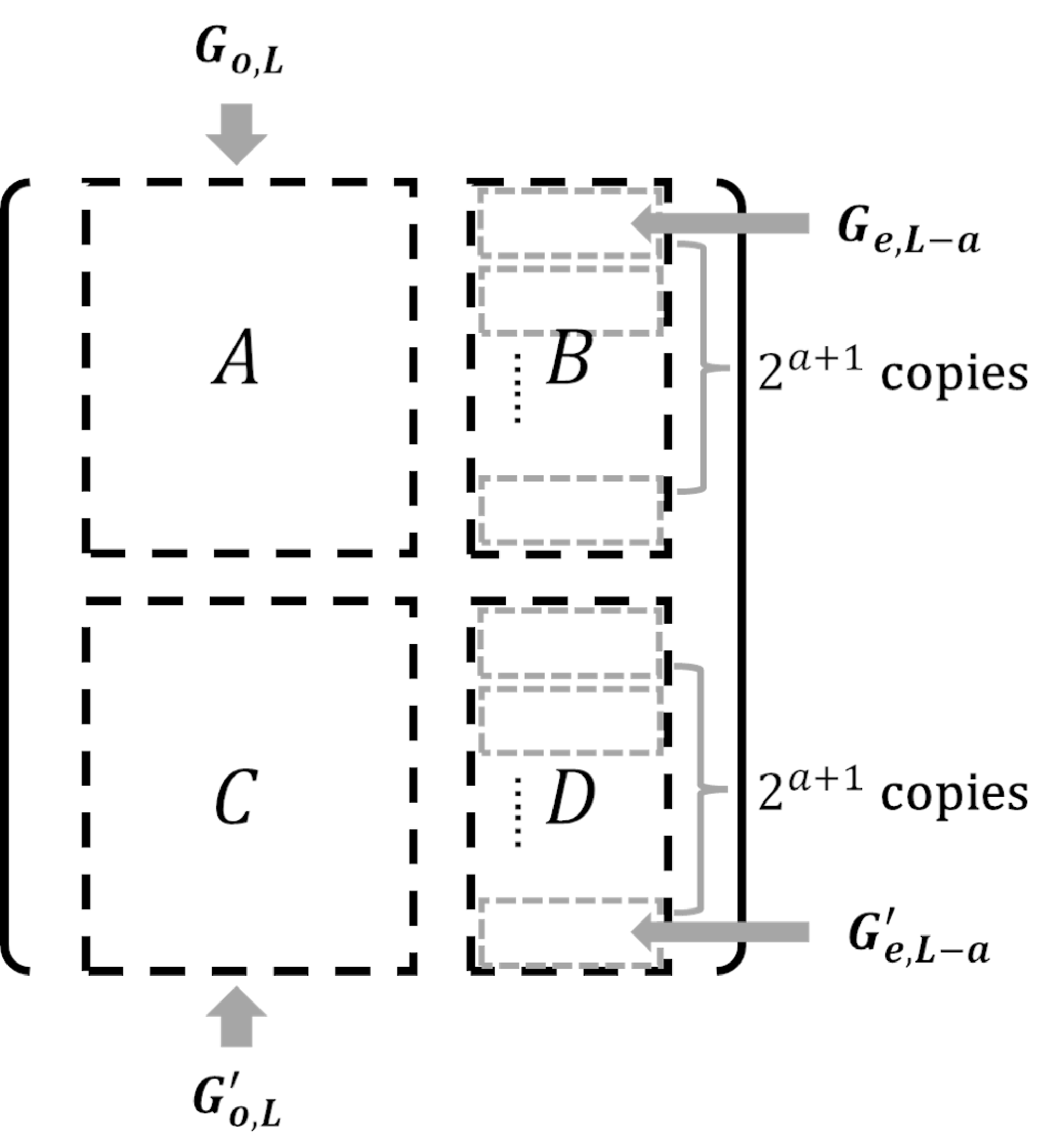}
	\caption{The structure of the check matrix $H$ of XZ-TGRE code. $H$ is composed of four submatrices $H_A$, $H_B$, $H_C$ and $H_D$, which are generated by Tanner graphs $G_{o,L}$, $G_{e,L-a}$, $G'_{o,L}$, and $G'_{e,L-a}$ shown in Fig. \ref{XZ-TGRE Tanner graph}, respectively. Notice that $2^{a+1}$ copies of the check matrix generated by $G_{e,L-a}$ and $G'_{e,L-a}$ form the complete $H_B$ and $H_D$, respectively.}
	\label{XZ-TGRE check matrix}
\end{figure}

\begin{figure*}
	\centering
	\includegraphics[width=0.8\textwidth]{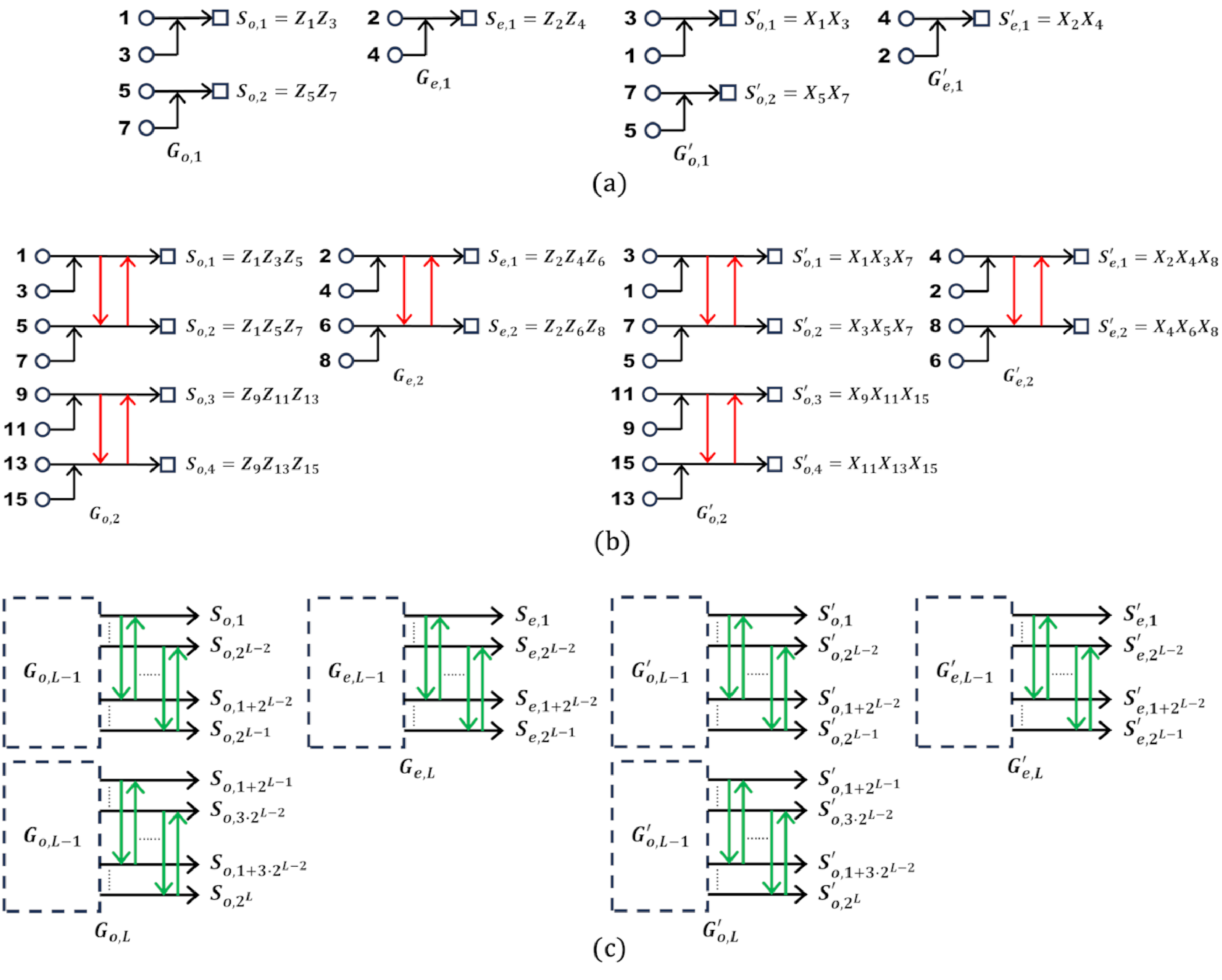}
	\caption{The recursive expansion of the Tanner graph of XZ-TGRE code. The subscript ``o'' of graph $G_o$ means the serial number of variable nodes in $G_o$ are odds. The subscript``e" of graph $G_e$ means the serial number of variable nodes in $G_e$ are evens. Each $G_o$ and $G_e$ has a corresponding graph denoted by $G'_o$ and $G'_e$ which are obtained by exchanging the number of variable nodes, respectively. The serial number of variable nodes in graph $G'$ is obtained by exchanging the serial number $b$ and $b+1$ in graph $G$, where $b=1+2^{2m}, m = 0, 1, 2, ...$ for $G'_o$ and $G_o$, or $b=2+2^{2m}, m = 0, 1, 2, ...$ for $G'_e$ and $G_e$. $G$s are used to generate Z type stabilizers, while $G'$s are used to generate X type stabilizers. The stabilizers generated by $G_o$, $G_e$, $G'_o$ and $G'_e$ are denoted by $S_o$, $S_e$, $S'_o$ and $S'_e$, respectively. (a) The primal Tanner graphs $G_{o,1}$, $G_{e,1}$, $G'_{o,1}$ and $G'_{e,1}$ used to expansion. (b) The expanded Tanner graphs $G_{o,2}$, $G_{e,2}$, $G'_{o,2}$ and $G'_{e,2}$ by the expansion of $G_{o,1}$, $G_{e,1}$, $G'_{o,1}$ and $G'_{e,1}$. (c) The expanded Tanner graphs $G_{o,L}$, $G_{e,L}$, $G'_{o,L}$ and $G'_{e,L}$ by the expansion of $G_{o,L-1}$, $G_{e,L-1}$, $G'_{o,L-1}$ and $G'_{e,L-1}$.}
	\label{XZ-TGRE Tanner graph}
\end{figure*}

Following the idea of recursively expanding Tanner graph, we propose XZ-TGRE code which can correct all three Pauli errors. The check matrix $H$ of XZ-TGRE is composed of four submatrices $H_A$, $H_B$, $H_C$ and $H_D$. As shown in Fig. \ref{XZ-TGRE check matrix}, each of them is obtained by an recursively expanded Tanner graph. The expansion method is shown in Fig. \ref{XZ-TGRE Tanner graph}. According to Fig. \ref{XZ-TGRE check matrix} and Fig. \ref{XZ-TGRE Tanner graph}, the code length is $N=2^{L-a}+2^{L+1}$. Proposition \ref{independence and commutativity of stabilizers of XZ-TGRE} 
states that the rows of the matrix composed of the submatrices $H_A$, $H_B$, $H_C$ and $H_D$ do form a set of stabilizer generators. It's obvious that the stabilizers corresponding to check matrix $H$ can be divided to two types -- Z type, which is denoted by $\{S_{i}\}$ and  generated by $G_{o,L}$ and $G_{e,L-a}$, contains only Pauli $Z$ operator, and X type, which is denoted by $\{S'_{j}\}$ and is generated by $G'_{o,L}$ and $G'_{e,L-a}$, contains only Pauli $X$ operator. 

\begin{proposition}
	\label{independence and commutativity of stabilizers of XZ-TGRE}
	The rows of $H$ are independent and commute with each other. 
\end{proposition}
\begin{proof}
According to the code construction of XZ-TGRE code in Fig. \ref{The structure of Hl} and Fig. \ref{XZ-TGRE Tanner graph} 
there are two types of stabilizers of XZ-TGRE code -- X type and Z type, which only contain Pauli $X$ and Pauli $Z$ operators, respectively. We only need to prove that all Z type stabilizers $\{S_i\}$ are independent with each other (since they must independent with X type) and commute with all X type stabilizers $\{S'_j\}$, then the independence and commutativity of all X type can be proved in the same way.

First, we prove all Z type stabilizers are independent with each other. Notice that each Z type stabilizer $S_i$ has a unique odd subscript $4i-1$. Hence, each $S_i$ cannot be generated by other Z type stabilizers.

Second, we prove all Z type stabilizers $\{S_i\}$ commute with all X type stabilizers $\{S'_j\}$. Notice that the serial number of variable nodes in the Tanner graph $G'$ is obtained by exchanging the serial number $b$ and $b+1$ in the Tanner graph $G$, where $b=1+2^{2m}, m = 0, 1, 2, ...$ for $G'_o$ and $G_o$, or $b=2+2^{2m}, m = 0, 1, 2, ...$ for $G'_e$ and $G_e$. Hence, by performing this number exchange and replacing the Pauli $Z$ by Pauli $X$ in the expression of the Z type stabilizers $\{S_i\}$, they will be transformed to the X type stabilizers $\{S'_j\}$. 

For a $S_i$, $\{S'_j\}$ can be divided into two class - Class 1 only contains the one $S'_i$ which can be obtained by the above transformation rules, and Class 2 contains the others.  According to the recursive expansion shown in Fig. 3 in the main text, the expression of $S'_i$ has two subscripts which that of $S_i$ also has, and the expression of $S'_j$ which belongs to Class 2 might have two or zero subscripts which that of $S_i$ also has. In any case, $S_i$ commute with $S'_j$. Hence, all Z type stabilizers $\{S_i\}$ commute with all X type stabilizers $\{S'_j\}$. 

The proof is completed.
	
\end{proof}

The parameter $a$ in Fig. \ref{XZ-TGRE check matrix} is used to control coding rate and code distance. According to Fig. \ref{XZ-TGRE check matrix} and Fig. \ref{XZ-TGRE Tanner graph}, the coding rate is $r = (2^{L-a})/(2^{L-a}+2^{L+1}) = 1/(1+2^{a+1})$. The relationship between the code distance and $a$ is given in Proposition \ref{code distance}. In the proof of Proposition \ref{code distance}, we also give out a form of the expression of XZ-TGRE code's logical operators.

\begin{table}
	\begin{center}
		\caption{The expression of the stabilizers and logical operators of XZ-TGRE code with code length $N=20$. The expression generated by $G_{o,3}$ and $G_{e,2}$ form the complete expression of stabilizers ${S_i}$ which are Z type. The expression generated by $G'_{o,3}$ and $G'_{e,2}$ form the complete expression of stabilizers ${S'_i}$ which are X type.}

		\begin{tabular}{c|c|c}
			\textbf{Stabilizers} & \ \makecell{\textbf{Expressions}\\ \textbf{generated by $G_{o,3}$}} & \ \makecell{\textbf{Expressions}\\ \textbf{generated by $G_{e,2}$}} \\
			\hline
			$S_{1}$ & $Z_{1}Z_{3}Z_{5}Z_{9}$ &  $Z_{2}Z_{4}Z_{6}$\\
			\hline
			$S_{2}$ & $Z_{1}Z_{5}Z_{7}Z_{13}$ & $Z_{2}Z_{6}Z_{8}$\\
			\hline
			$S_{3}$ & $Z_{1}Z_{9}Z_{11}Z_{13}$ & $Z_{2}Z_{4}Z_{6}$\\
			\hline
			$S_{4}$ & $Z_{5}Z_{9}Z_{13}Z_{15}$ & $Z_{2}Z_{6}Z_{8}$\\
			\hline
			$S_{5}$ & $Z_{17}Z_{19}Z_{21}Z_{25}$ & $Z_{2}Z_{4}Z_{6}$\\
			\hline
			$S_{6}$ & $Z_{17}Z_{21}Z_{23}Z_{29}$ & $Z_{2}Z_{6}Z_{8}$\\
			\hline
			$S_{7}$ & $Z_{17}Z_{25}Z_{27}Z_{29}$ & $Z_{2}Z_{4}Z_{6}$\\
			\hline
			$S_{8}$ & $Z_{21}Z_{25}Z_{29}Z_{31}$ & $Z_{2}Z_{6}Z_{8}$\\
			\hline
			
			\textbf{Stabilizers} & \ \makecell{\textbf{Expressions}\\ \textbf{generated by $G'_{o,3}$}} & \ \makecell{\textbf{Expressions}\\ \textbf{generated by $G'_{e,2}$}} \\
			\hline
			$S'_{1}$ & $X_{1}X_{3}X_{7}X_{11}$ & $X_{2}X_{4}X_{8}$\\
			\hline
			$S'_{2}$ & $X_{3}X_{5}X_{7}X_{11}$ & $X_{4}X_{6}X_{8}$\\
			\hline
			$S'_{3}$ & $X_{3}X_{9}X_{11}X_{15}$ & $X_{2}X_{4}X_{8}$\\
			\hline
			$S'_{4}$ & $X_{7}X_{11}X_{13}X_{15}$ & $X_{4}X_{6}X_{8}$\\
			\hline
			$S'_{5}$ & $X_{17}X_{19}X_{23}X_{27}$ & $X_{2}X_{4}X_{8}$\\
			\hline
			$S'_{6}$ & $X_{19}X_{21}X_{23}X_{31}$ & $X_{4}X_{6}X_{8}$\\
			\hline
			$S'_{7}$ & $X_{19}X_{25}X_{27}X_{31}$ & $X_{2}X_{4}X_{8}$\\
			\hline
			$S'_{8}$ & $X_{23}X_{27}X_{29}X_{31}$ & $X_{4}X_{6}X_{8}$\\
			\hline
			\makecell{\textbf{Logical} \\ \textbf{qubit} \\ \textbf{number}}& \ \makecell{\textbf{Logical $X$} \\ \textbf{operators}} & \ \makecell{\textbf{Logical $Z$} \\ \textbf{operators}} \\
			\hline
			\rule{0pt}{10pt}1 & $\bar{X}_1 = X_{2}X_{4}X_{8}$ & $\bar{Z}_1 = Z_{2}Z_{1}Z_{9}Z_{17}Z_{25}$ \\
			\hline
			\rule{0pt}{10pt}2 & $\bar{X}_2 = X_{4}X_{6}X_{8}$ & $\bar{Z}_2 = Z_{6}Z_{5}Z_{13}Z_{21}Z_{29}$ \\
			\hline
			\rule{0pt}{10pt}3 & $\bar{X}_3 = X_{4}X_{3}X_{11}X_{19}X_{27}$ & $\bar{Z}_3 = Z_{2}Z_{4}Z_{6}$ \\
			\hline
			\rule{0pt}{10pt}4 & $\bar{X}_4 = X_{8}X_{7}X_{15}X_{23}X_{31}$ & $\bar{Z}_4 = Z_{2}Z_{6}Z_{8}$ \\
			\hline
		\end{tabular}
		\label{table_XZ-TGRE}
	\end{center}
	
\end{table}

\begin{proposition}
	\label{code distance}
	For a given parameter $a$, the code distance of XZ-TGRE code $d \leq 1+2^{a+1}$. The equality holds when $L$ reaches $L \sim 1+2^{a+1}+a$
\end{proposition}
\begin{proof}
To prove Proposition \ref{code distance}, we need to give out one form of the logical operators at first.

Take the XZ-TGRE code with $N=20$ ($L=3$ and $a=1$) for an example. The expression of logical operators is shown in Table \ref{table_XZ-TGRE}. All logical operators can be divided to two types -- the subscripts of the expression of Type 1 are all evens, and those of Type 2 are one even combined with several odds.

Type 1 logical operators are obtained from the submatrices corresponding to Tanner graph $G_{e,L-a}$ and $G'_{e,L-a}$.  There is a unique even number in the expressions of each $\bar{X}_i$ among all $\bar{X}$ belong to Type 1 (in the example, those numbers are 2 and 6) and a unique even number in each $\bar{Z}_i$ among all $\bar{Z}$ belong to Type 1 (in the example, those numbers are 4 and 8), according to which we can write out Type 2 logical operators. Let's take $\bar{X}_1$ in Table \ref{table_XZ-TGRE} for an instance. In the expression of $\bar{X}_1$, there is a unique number 2 which only occurs in $\bar{X}_1$ but not in other $\bar{X}_i$ belong to Type 1. If the expression of $\bar{Z}_1$ contains $Z_2$ but no other evens occurs in $\bar{X}_1$ and other $\bar{X}_i$ belong to Type 1, then $\bar{Z}_1$ will naturally anti-commute with $\bar{X}_1$ and commute with other $\bar{X}_i$ belong to Type 1. What's more, to ensure $\bar{Z}_1$ commutes with all X-type stabilizers, we must add some Pauli $Z$s with odd subscripts to the expression of $\bar{Z}_1$. We can easily find there is a unique odd in each stabilizer which contains $X_2$, which are 1, 9, 17 and 25. Adding Pauli $Z$s with these odds to the expression of $\bar{Z}_1$ will ensure $\bar{Z}_1$ commutes with all stabilizers. Now we get a logical $Z$ operator belongs to Type 2.

It's easy to prove that Type 1 and Type 2 logical operators don't belong to the stabilizer group and commute with all stabilizers, and that all logical operators commute with each other except the ones perform on the same logical qubit, which anti-commutes with each other. Also, it's obvious that all logical operators are independent with each other.

If parameter $a$ is fixed, the weight of Type 1 logical operators will naturally increase with the code length. However, the weight of Type 2 logical operators won't change, because that the number of the copies in the submatrices $H_B$ and $H_D$ is fixed to $2^{a+1}$. According to the above method to write out Type 2 logical operators by Type 1, the weight of Type 2 logical operators will be fixed to $1+2^{a+1}$. 
	
\end{proof}

Proposition \ref{the minimum weight of Type 1 and 2} gives some information about the minimum weight of Type 1 and Type 2 logical operators. 

\begin{proposition}
	\label{the minimum weight of Type 1 and 2}
	Type 2 logical operators are already in the minimum-weight form. The only way to reduce the weight of Type 1 logical operators is multiplying one by several others instead of stabilizers.
\end{proposition}
\begin{proof}
	
	For a Type 1 logical operator $\bar{L}_i$, if one tries to reduce its weight with stabilizers through eliminating several Pauli operators with even subscripts, the most efficient way is to use the stabilizers which contains $\bar{L}_i$, which will actually in turn introduce more Pauli operators with odd subscripts. Hence, stabilizers cannot be used to reduce the weight of Type 1 logical operators. The only way to reduce the weight of Type 1 logical operators is multiplying one by several others.
	
	For a Type 2 logical operator $\bar{L}_j$, its expression can be written as 
	
	\begin{equation}
		\begin{aligned}
			\bar{L}_j  = & P_{4j-2}P_{4j-3}P_{4j-3+2^{L-a+1}}P_{4j-3+2\cdot2^{L-a+1}} \\
			& P_{4j-3+3\cdot2^{L-a+1}}...P_{4j-3+(2^{a+1}-1)2^{L-a+1}}
		\end{aligned}
		\label{L_j}
	\end{equation}
	where $P \in \{ X, Z \}$.
	
	Without loss of generality, we only need to prove Proposition \ref{the minimum weight of Type 1 and 2} holds for Type 2 logical $Z$ operator $\bar{Z}_j$. Then, the proof for Type 2 $\bar{X}_j$ will be obvious. 
	
	The expression of $\bar{Z}_j$ is
	
	\begin{equation}
		\begin{aligned}
			\bar{Z}_j  = & Z_{4j-2}Z_{4j-3}Z_{4j-3+2^{L-a+1}}Z_{4j-3+2\cdot2^{L-a+1}} \\
			& Z_{4j-3+3\cdot2^{L-a+1}}...Z_{4j-3+(2^{a+1}-1)2^{L-a+1}}\\
			=& Z_{jI}Z_{jII}Z_{jIII}
		\end{aligned}
		\label{z_j}
	\end{equation}
	where $Z_{jI} = Z_{4j-2}$, $Z_{jII} =Z_{4j-3}Z_{4j-3+2^{L-a+1}}...Z_{4j-3+(2^{a}-1)\cdot2^{L-a+1}}$, and $Z_{jIII}=Z_{4j-3+2^{a}\cdot2^{L-a+1}}...Z_{4j-3+(2^{a+1}-1)2^{L-a+1}}$.
	
	Notice that only the stabilizers ${S_e}$ can be used to reduce the weight of the part $Z_{jI}$, and only the stabilizers ${S_o}$ can be used to reduce the weight of the part $Z_{jII}$ and $Z_{jIII}$. We only need to prove the weight of $Z_{jII}$, which is denoted by $wt_{II}$, cannot be reduced by ${S_o}$. The proof for $Z_{jI}$ and $Z_{jIII}$ is similar. 
	
	Using the same method of the proof of Proposition \ref{the minimum weight of logical X of Z-TGRE}, we can readily prove the minimum weight of elements in the group with generators $<S_{o,1},...,S_{o,2^L}>$, which is denoted by $wt_{min}(S_o)$, is $L+1$ (if $L$ is an odd) or $L$ (if $L$ is an even). Hence, $ wt_{min}(S_o) \geq L$. Besides, only when $L$ reaches $L \sim 1+2^{a+1}+a$, $a$ will be added by 1. According to this, $L \geq 2^{a+1} = 2wt_{II}$. That is to say, even there is an element $g$ in in the group with generators $<S_{o,1},...,S_{o,2^L}>$ contains $Z_{jII}$, since $wt(g)\geq2wt_{II}$, $g$ cannot be used to reduce $wt_{II}$. Hence, the weight of $Z_{jII}$ cannot be reduced, and the same to $Z_{jI}$ and $Z_{jIII}$. 
	
	In addition, for any $\bar{Z}_j$ and $\bar{Z}_m$ $(j \neq m)$, they do not have any subscript which is the same. Hence, the weight of $\bar{Z}_j$ cannot be reduced by any $\bar{Z}_m$ $(j \neq m)$. Therefore, the expression Eq. (\ref{z_j}) of $\bar{Z}_j$ is the minimum-weight form.
	
	The proof is completed.
\end{proof}

In this situation, if the minimum weight of Type 1 logical operators is less than the weight of the Type 2, the code distance is decided by the minimum weight of Type 1, or else, the code distance will be the weight of Type 2 logical operators. Comparing Type 1 logical operators with the logical $X$ operators of Z-TGRE code, we can find the relationship between the minimum weight of Type 1 and code length should be similar to the relationship between the minimum weight of the logical $X$ operators of Z-TGRE code and its code length, which is clarified in Proposition \ref{the minimum weight of logical X of Z-TGRE}. Besides, when the code distance reaches the weight of Type 2, $L-a \sim 1+2^{a+1}$, namely, $L \sim 1+2^{a+1}+a$.

For stabilizer codes, if code distance doesn't increase with code length, then the logical error rate cannot be lowered by increasing the code length. To ensure the code distance of XZ-TGRE code will increase with the code length, when $L$ reaches $L \sim 1+2^{a+1}+a$, we must raise the weight of Type 2 by increasing parameter $a$ through adding 1 to it, which in turn reduces the coding rate.

\begin{figure}[htbp]
	\centering
	\includegraphics[width=0.5\textwidth]{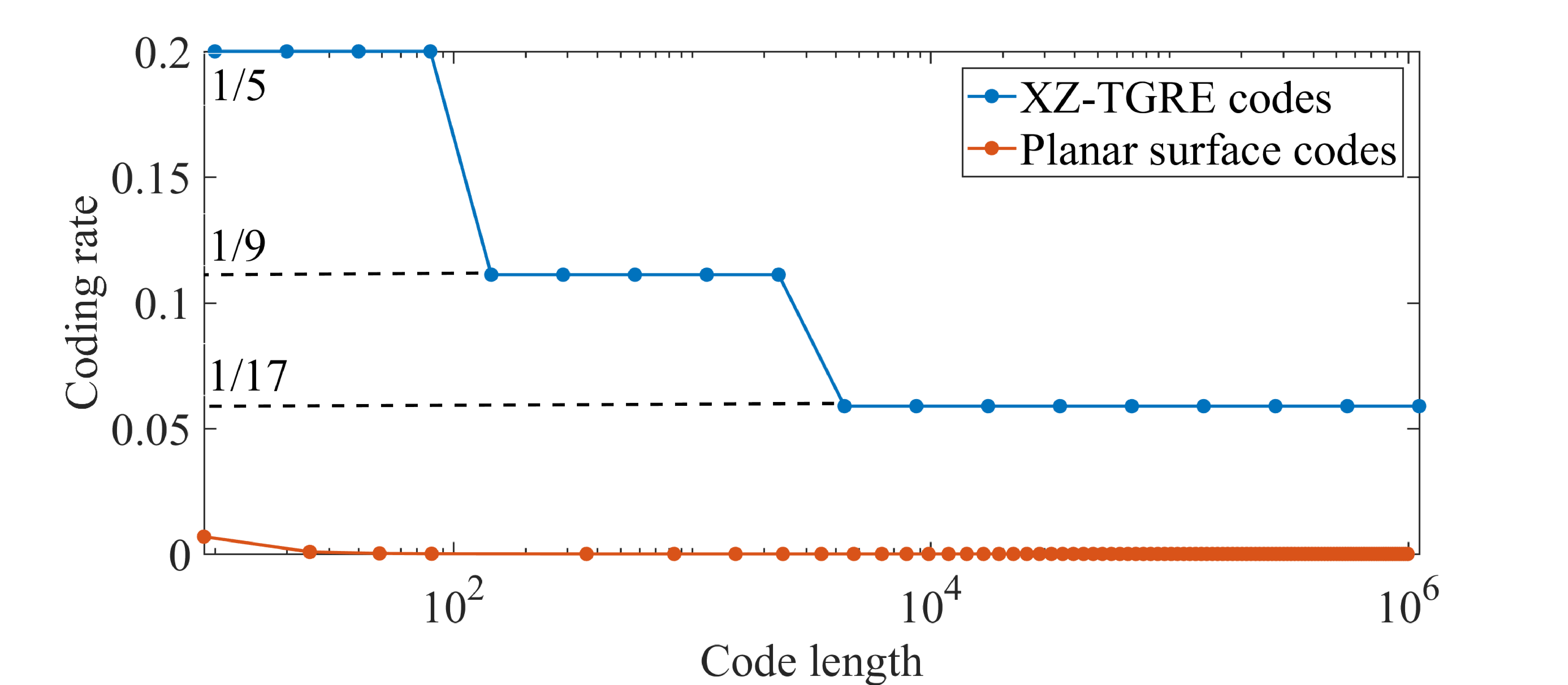}
	\caption{The coding rate of XZ-TGRE code and planar surface code.}
	\label{coding rate of XZ-TGRE}
\end{figure}

Table \ref{table_XZ-TGRE_distance} shows the code distance of XZ-TGRE code with different code length, which is determined by the Monte Carlo method in \cite{liang2024determining}. Fig. \ref{coding rate of XZ-TGRE} shows the practical coding rate of XZ-TGRE code. According to Fig. \ref{coding rate of XZ-TGRE}, though XZ-TGRE code has zero asymptotic coding rate, its practical coding rate tends to zero extremely slowly and much more slowly than surface code. In a fairly long code length range, the coding rate of XZ-TGRE is even higher than many stabilizer codes with constant coding rate, such as (3,4) hypergraph product code\cite{grospellier2021combining}, (5,6) hypergraph product code\cite{grospellier2018numerical} and  (4,5)-hyperbolic
surface code\cite{breuckmann2016constructions}. 

It should be noticed that the weight of the stabilizers of XZ-TGRE code scales as $O(\log N)$.

\begin{table}[htbp]
	\begin{center}
		\caption{The code distance $d$ and coding rate $r$ of XZ-TGRE code with code length $N$ from $20$ to $1152$. $L$ and $a$ is the subscript of Tanner graphs $G_{o,L}$, $G_{e,L}$, $G'_{o,L-a}$ and $G'_{e,L-a}$. $wt_{min}(\bar{X})$, $wt_{min}(\bar{Y})$ and $wt_{min}(\bar{Z})$ are the minimum weight of logical $X$, $Y$ and $Z$ operators, respectively.}		
		\begin{tabular}{c|c|c|c|c|c|c|c}
			\textbf{$L$} & \ \textbf{$a$} & \ $N$ & \ $wt_{min}(\bar{X})$ & \ $wt_{min}(\bar{Z})$ &\ $wt_{min}(\bar{Y})$ & \ $d$ & \ $r$ \\
			\hline
			3 & 1 & 20 & 2 & 2 & 3 & 2 & 1/5 \\
			\hline
			4 & 1 & 40 & 4 & 4 & 4 & 4 & 1/5 \\
			\hline
			5 & 1 & 80 & 4 & 4 & 5 & 4 & 1/5 \\
			\hline
			6 & 2 & 144 & 4 & 4 & 5 & 4 & 1/9 \\
			\hline
			7 & 2 & 288 & 6 & 6 & 6 & 6 & 1/9 \\
			\hline
			8 & 2 & 576 & 6 & 7 & 7 & 6 & 1/9 \\
			\hline
			9 & 2 & 1152 & 8 & 8 & 9 & 8 & 1/9 \\
			\hline
		\end{tabular}
		\label{table_XZ-TGRE_distance}
	\end{center}
	
\end{table}

\section{TGRE-HP code}
\label{TGRE-HP code}
Hypergraph product codes is a class of quantum LDPC codes proposed by Tillich and Zémor\cite{tillich2013quantum}. They are constructed by Cartesian product of two Tanner graphs $G_A$ and $G_B$. Assume $G_A$ corresponds to code A $[n_A,k_A,d_A]$ and $G_B$ corresponds to code B $[n_B,k_B,d_B]$. The code parameters of hypergraph product code constructed by them will be

\begin{equation}
	\label{code parameters}
	[n_An_B+(n_A-k_A)(n_B-k_B),k_Ak_B,min(d_A,d_B)]
\end{equation}
which has been proved in \cite{tillich2013quantum}. According to this, if one select code A and code B properly, it might be possible to obtain a hypergraph product code with high coding rate or large code distance. 

\begin{figure}
	\centering
	\includegraphics[width=0.4\textwidth]{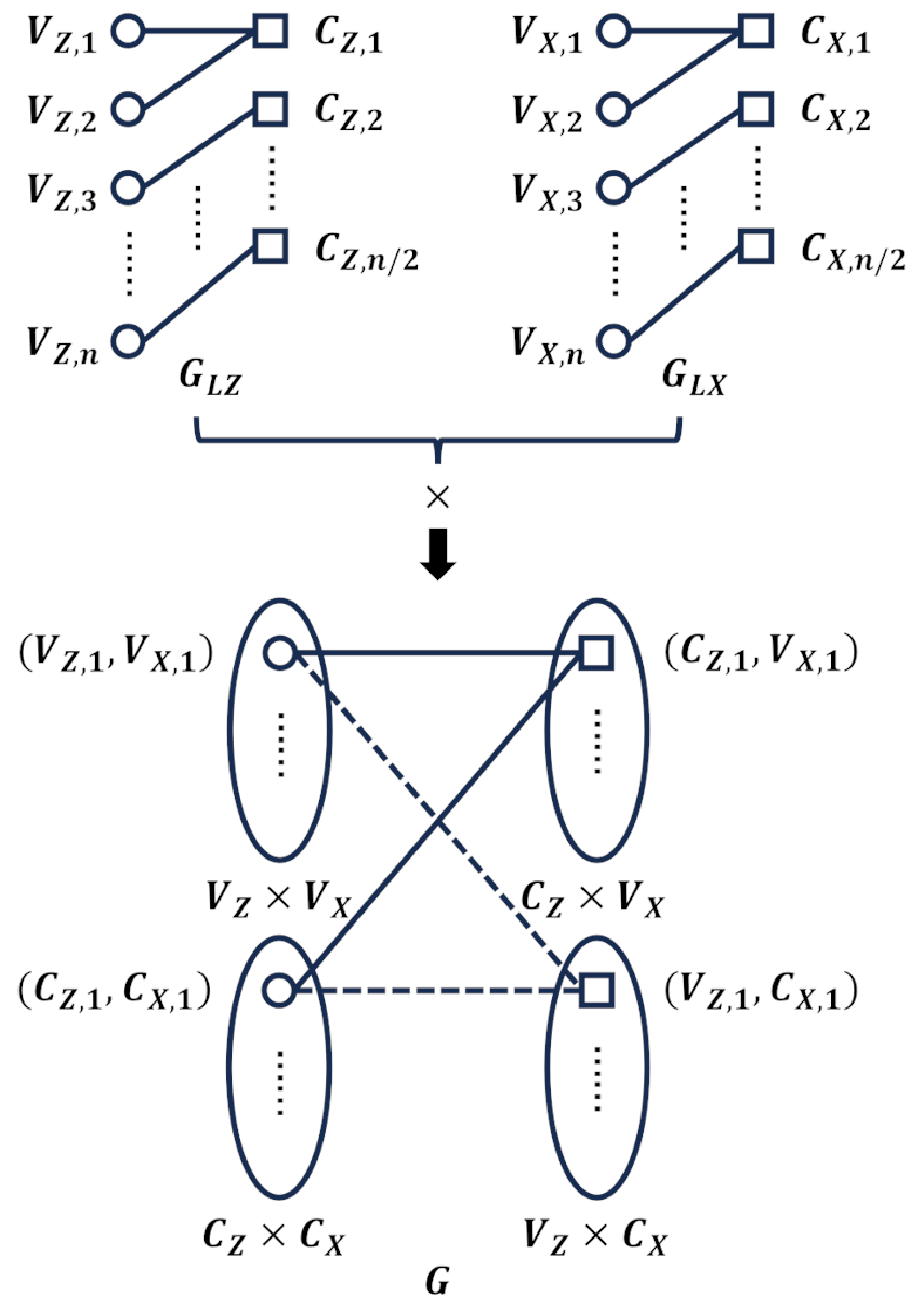}
	\caption{The Tanner graph of TGRE-HP code. The "$\times$" means Cartesian product of two sets. $G_{LZ}$ and $G_{LX}$ are the Tanner graphs of the Z-TGRE code and X-TGRE code with code length $n=2^L$, respectively. $G$ is the Tanner graph of the TGRE-HP code with code length $N = \frac{5}{4} n^2$.}
	\label{graph product}
\end{figure}

As shown in Fig. \ref{graph product}, the TGRE-HP code with code length $N$ are constructed by a Z-TGRE code with code length $n$ and a X-TGRE code with code length $n$. According to Eq. (\ref{code parameters}), the code parameters of this TGRE-HP code are $[\frac{5}{4} n^2, \frac{1}{4}n^2, d]$, where $d$ is the code distance of the Z-TGRE code (and also the code distance of the X-TGRE code). Hence, with the growth of the code length of the Z-TGRE code and X-TGRE code used to construct TGRE-HP code, the coding rate is always 0.2, and the code distance scales as  $O(\log n) = O(\log \sqrt{N})$. Besides, according to graph $G$ in Fig. 2, the weight of the stabilizers will also increase with the growth of code length at a rate of $O(\sqrt{N})$, or more accurately, $\frac{3}{\sqrt{5}}\sqrt{N}$.

As for the logical operators, they can be given out through the standard form of the check matrix\cite{nielsen2001quantum}.

\section{Simulation results}
\label{simulation results}

\begin{figure}[htbp]
	\centering
	\includegraphics[width=0.5\textwidth]{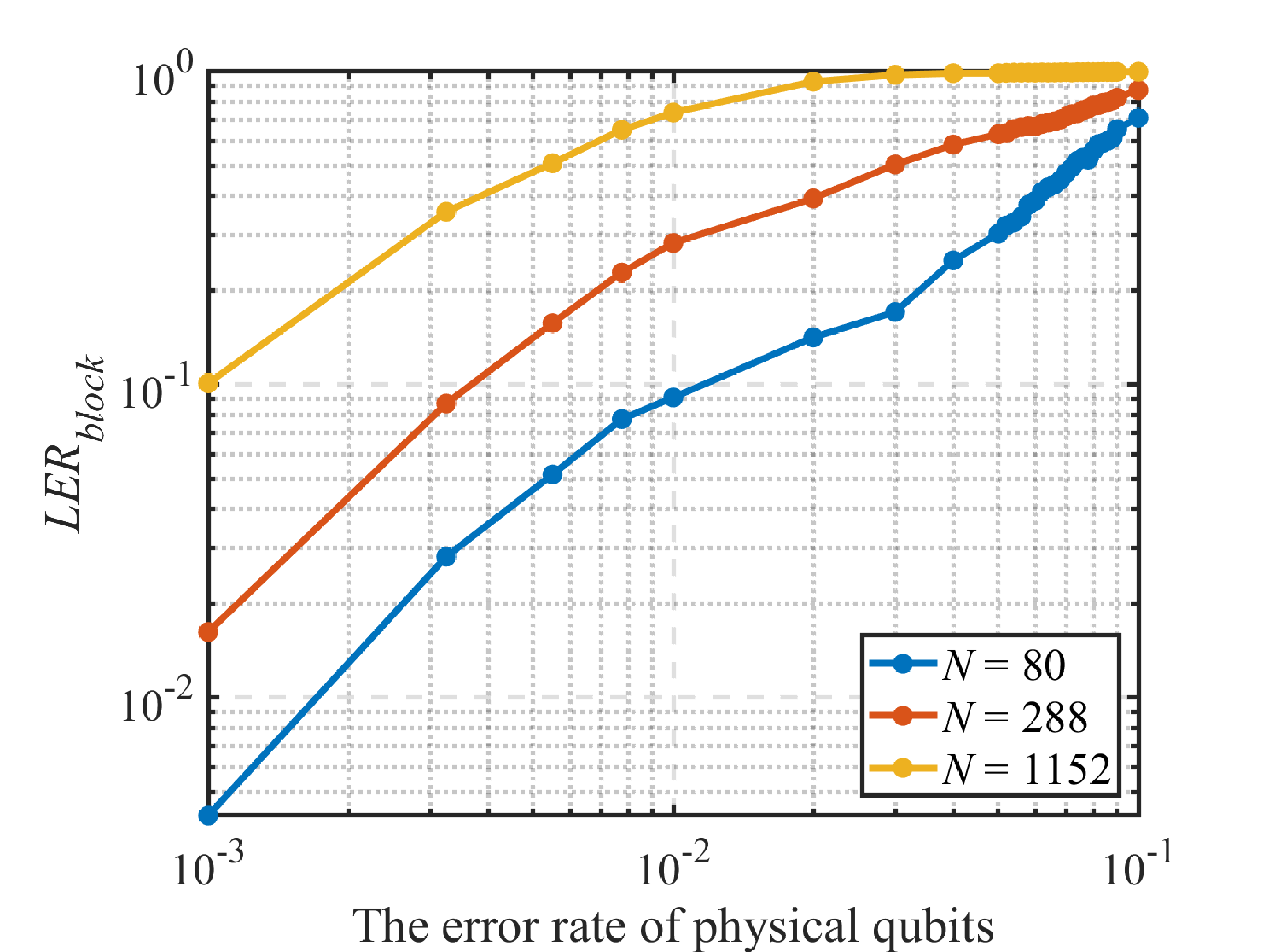}
	\caption{The $LER_{block}$ of XZ-TGRE code in depolarizing noise channel.}
	\label{ler_block}
\end{figure}

\begin{figure}[htbp]
	
	\centering
	\includegraphics[width=0.5\textwidth]{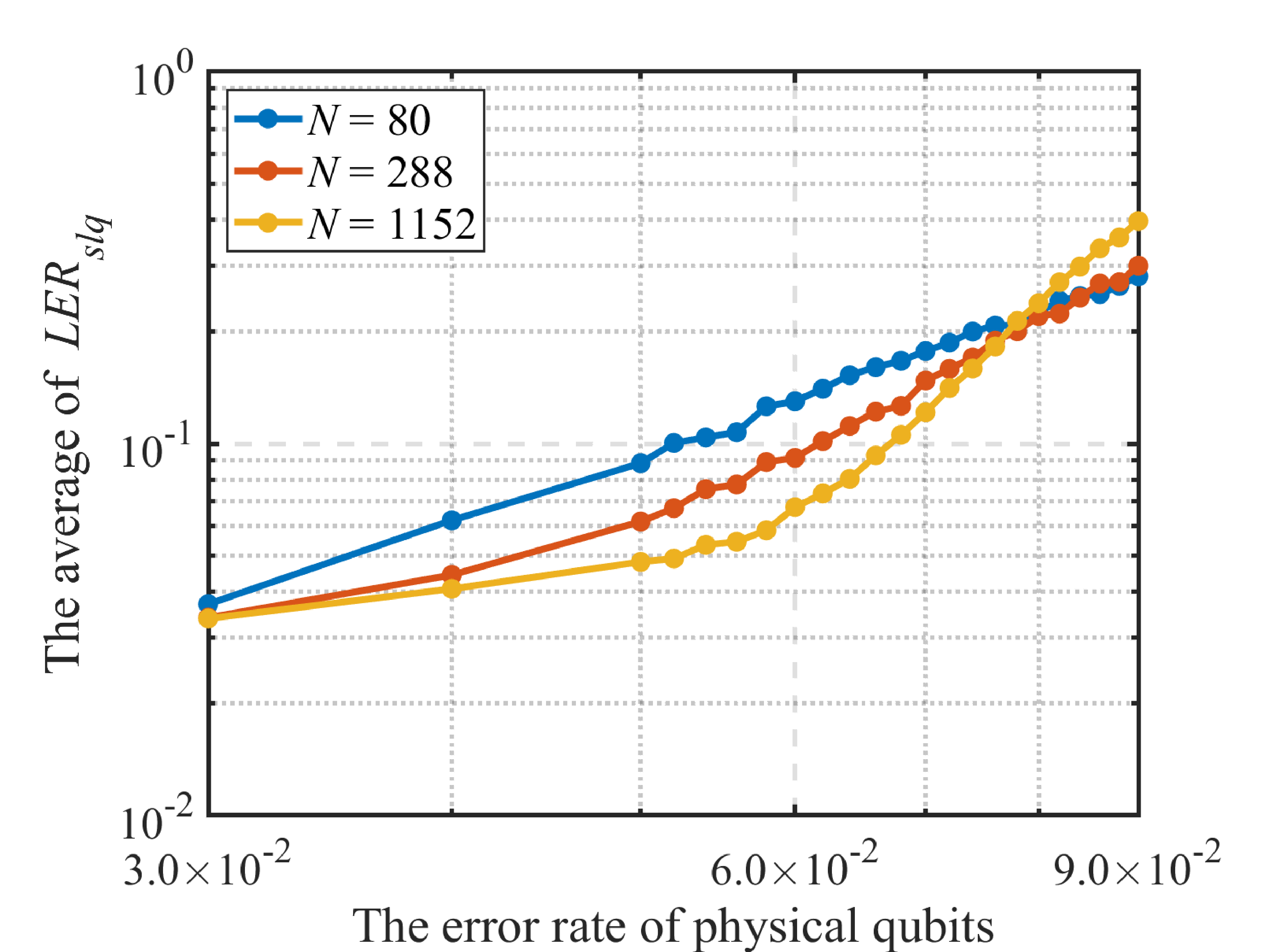}
	\caption{The average of $LER_{slq}$ of XZ-TGRE code in depolarizing noise channel.}
	\label{ler_slq}
\end{figure}

\begin{figure}[htbp]
	\centering
	\includegraphics[width=0.5\textwidth]{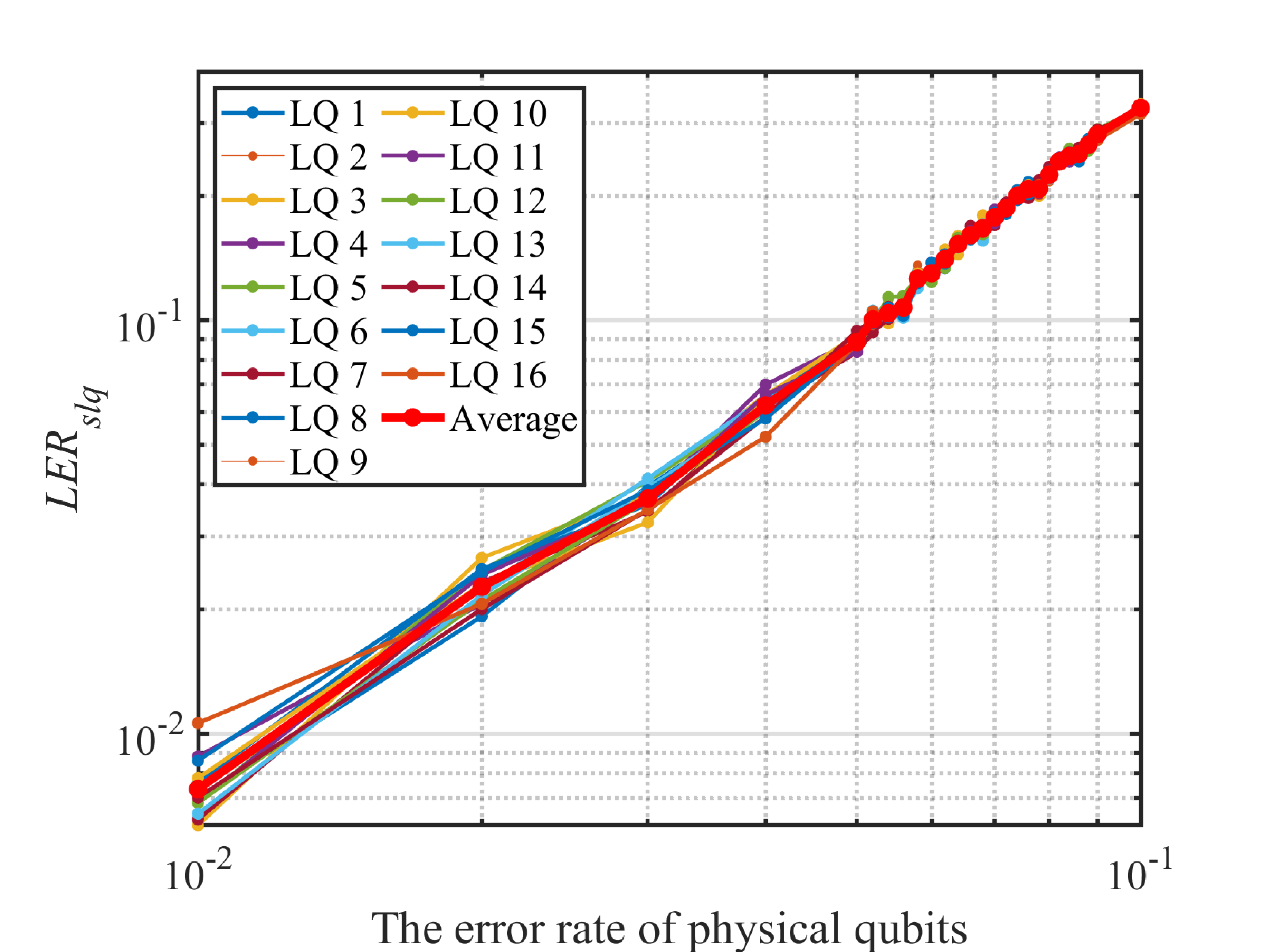}
	\caption{The error rate of every logical qubit of XZ-TGRE code when code length $N=80$ in depolarizing noise channel. LQ means logical qubit.}
	\label{ler_slq of every logical qubit}
\end{figure}

\begin{figure}[htb]
	
	\centering
	\includegraphics[width=0.5\textwidth]{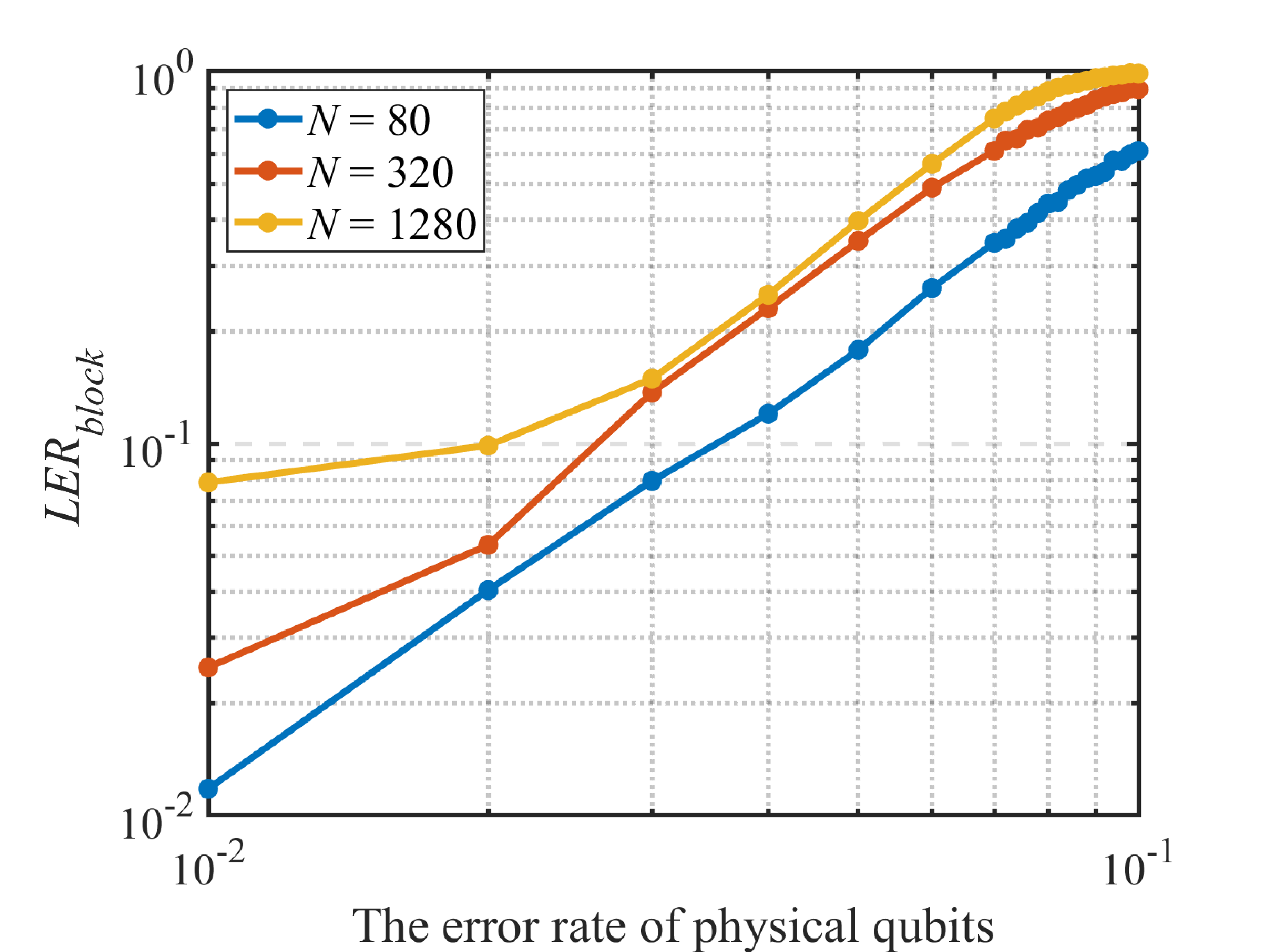}
	\caption{The $LER_{block}$ of TGRE-HP code in depolarizing noise channel.}
	\label{ler_block_hp}
\end{figure}

\begin{figure}[htb]
	\centering
	\includegraphics[width=0.5\textwidth]{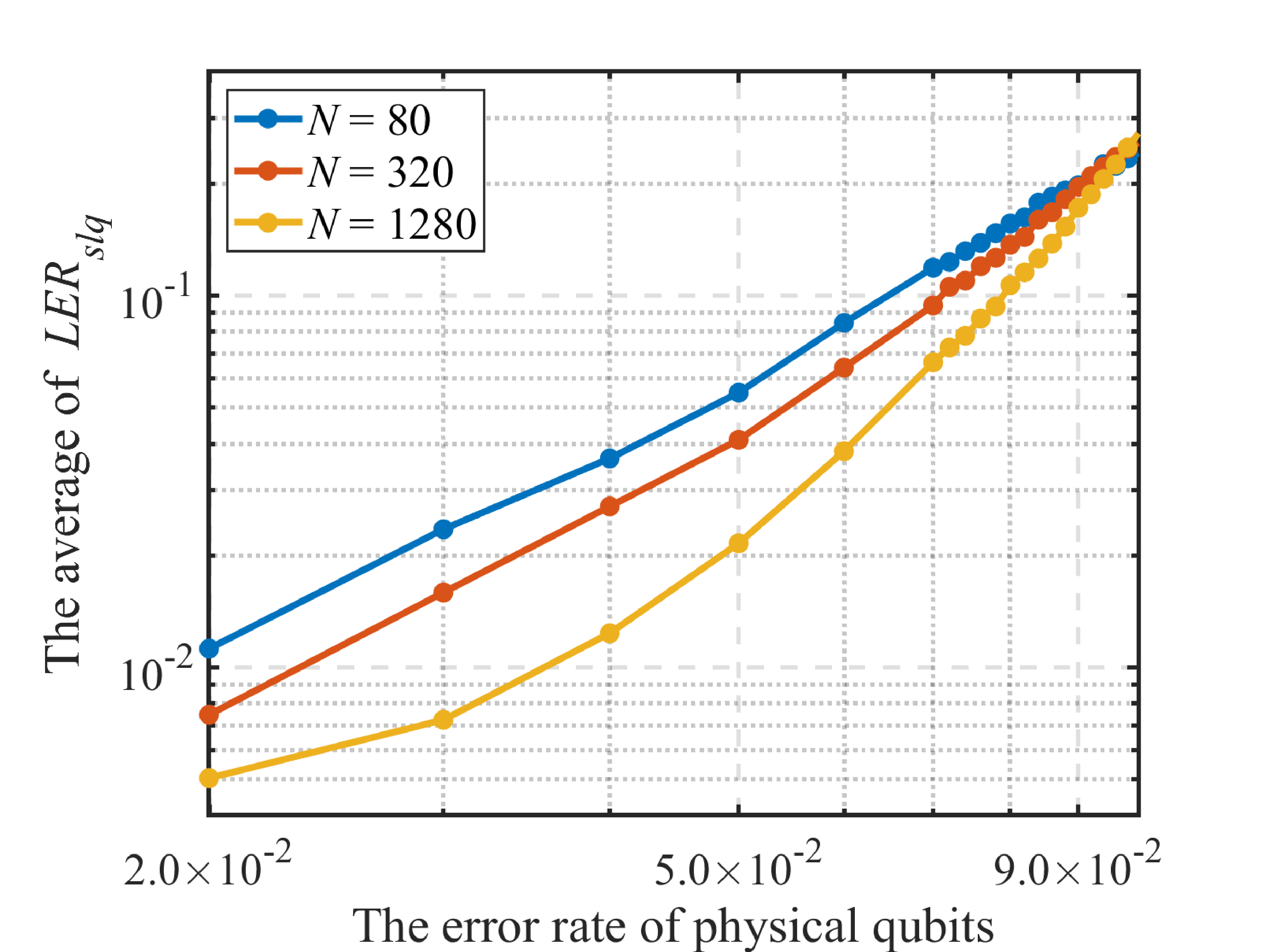}
	\caption{The average of $LER_{slq}$ of TGRE-HP code in depolarizing noise channel.}
	\label{ler_slq_hp}
\end{figure}

\begin{figure}[htb]
	\centering
	\includegraphics[width=0.5\textwidth]{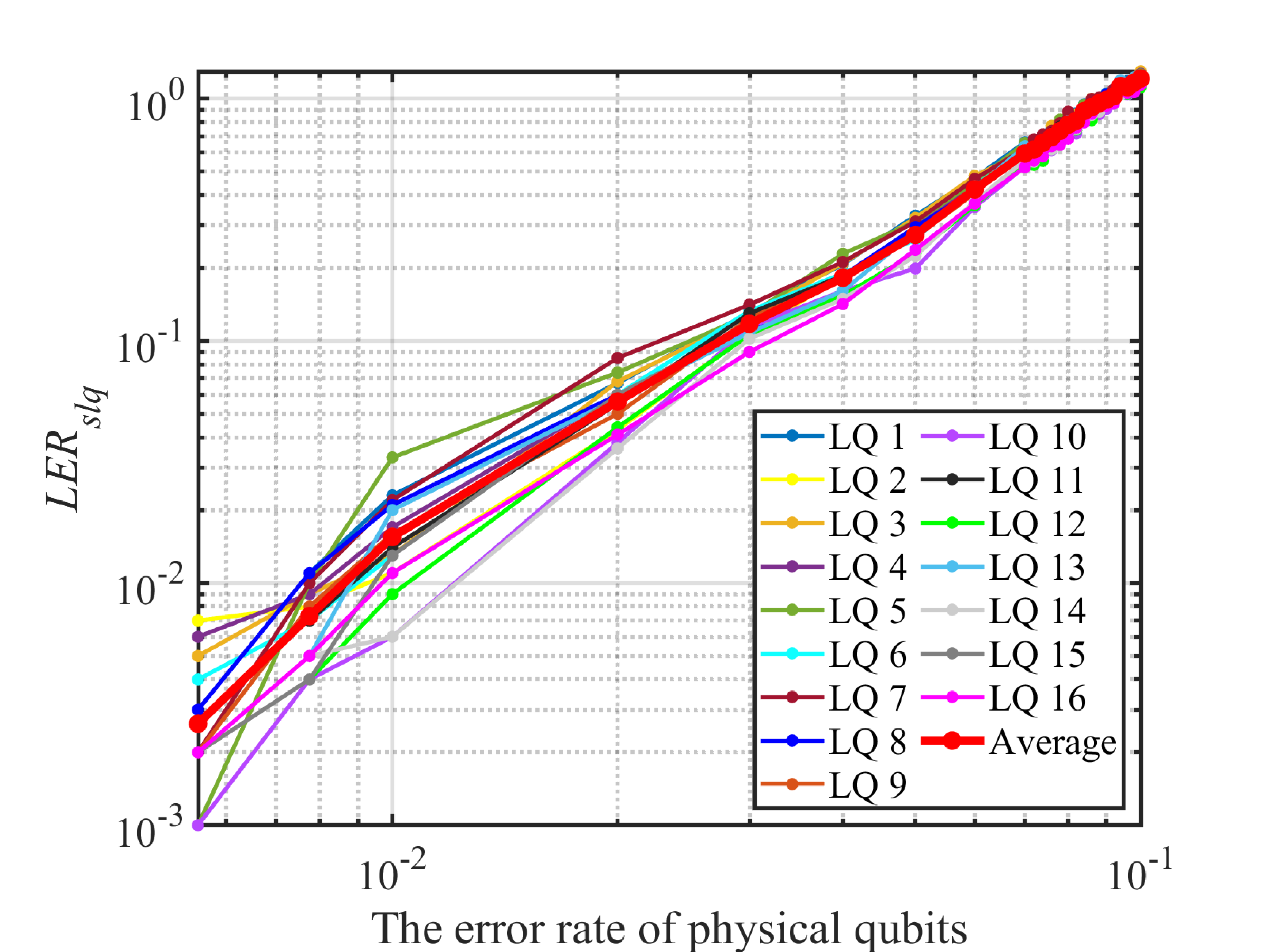}
	\caption{The $LER_{slq}$ every logical qubit of TGRE-HP code when code length $N=80$ in depolarizing noise channel. LQ means logical qubit.}
	\label{ler_slq of every logical qubit_hp}
\end{figure}

Fig. \ref{ler_block} to Fig. \ref{ler_slq of every logical qubit} show the error correcting simulation results of XZ-TGRE code in depolarizing noise channel, and Fig. \ref{ler_block_hp} to Fig. \ref{ler_slq of every logical qubit_hp} show the error correcting simulation results of TGRE-HP code in depolarizing noise channel. All the simulation is performed under the assumption that the encoding operations, stabilizer measurement and recovery operations are all perfect, and is realized with the FDBP decoding algorithm proposed in our previous work\cite{yi2023improved}, whose decoding accuracy is higher than traditional BP over GF(2). The logical error rate of the whole code block is denoted by $LER_{block}$, and the logical error rate of single logical qubit is denoted by $LER_{slq}$. 

Fig. \ref{ler_slq of every logical qubit} and Fig. \ref{ler_slq of every logical qubit_hp} show the $LER_{slq}$ of each logical qubit when the code length is 80. According to these two figures, for XZ-TGRE code and TGRE-HP code, the average of $LER_{slq}$ can represent the $LER_{slq}$ of each logical qubit well. According to Fig. \ref{ler_slq} and Fig. \ref{ler_slq_hp}, the code capacity noise threshold of XZ-TGRE code is around 0.078, and that of TGRE-HP code is around 0.096. It is worth noting that the decoding accuracy of FDBP will be influenced by the weight of stabilizers. In general, greater weight of stabilizers will lead to lower decoding accuracy of EDBP. This is because the larger the weight of stabilizers is, the more the short loops might be in the Tanner graphs, which do harms to the decoding accuracy of FDBP. Notice that the weight of the stabilizers of XZ-TGRE code and TGRE-HP code increase with code length at a rate of $O(\log N)$ and $O(\sqrt{N})$, respectively. This means that the longer the code length is, the lower the decoding accuracy will be. Hence, we have reasons to believe the actual logical error rate in Fig. \ref{ler_slq} should be lower and the actual code capacity noise threshold should be higher.

\section {Conclusion}
\label{conclusion}

In this article, we propose XZ-TGRE code and TGRE-HP code. To ensure the code distance will increase with code length, XZ-TGRE code sacrifices the coding rate, which leads to zero asymptotic coding rate. Even so, the coding rate of XZ-TGRE code tends to zero extremely slowly. In a considerable range of code length, the coding rate of XZ-TGRE code is larger than some QLDPC with constant coding rate. The coding rate of TGRE-HP code is the constant 0.2 which is the highest among the existing quantum stabilizer codes to our best knowledge. The code distance of XZ-TGRE code scales as $O(log(N))$, and that of TGRE-HP code scales as $O(\log \sqrt{N})$, where $N$ is the code length. Besides, the weight of the stabilizers of XZ-TGRE code scales as $O(\log N)$, and that of the TGRE-HP code scales as $O(\sqrt{N})$.

We use FDBP decoding algorithm to perform the error correcting simulation and find the code capacity noise threshold of XZ-TGRE code and TGRE-HP code in depolarizing channel is around 0.078 and 0.096, respectively. Due to the limit of FDBP, the actual code capacity noise threshold might be even higher than the simulation results. Our work shows that the idea of recursively expanding Tanner graph might have potential to construct quantum stabilizer codes with better performance.

\newpage
\section*{End Notes}
\subsection*{Acknowledgements}
 This work was supported by the Colleges and Universities Stable Support Project of Shenzhen, China (No.GXWD20220817164856008), the Colleges and Universities Stable Support Project of Shenzhen, China (No.GXWD20220811170225001), the Colleges and Universities Stable Support Project of Shenzhen, China, (No. GXWD20220811170225001) and Harbin Institute of Technology, Shenzhen - SpinQ quantum information Joint Research Center Project (No.HITSZ20230111).

\section*{Data Availability}
The data that support the findings of this study are
available from the corresponding author upon reasonable
request.

\bibliography{reference}

\end{document}